\documentclass[aps,prd,nofootinbib,twocolumn,superscriptaddress,preprintnumbers]{revtex4}

\usepackage{amssymb}
\usepackage{amsmath}
\usepackage{epsfig}
\usepackage{hyperref}
\usepackage{breakurl}

\usepackage{undertilde}
\makeatletter
\def\simgt{\mathrel{\lower2.5pt\vbox{\lineskip=0pt\baselineskip=0pt
           \hbox{$>$}\hbox{$\sim$}}}}
\def\simlt{\mathrel{\lower2.5pt\vbox{\lineskip=0pt\baselineskip=0pt
           \hbox{$<$}\hbox{$\sim$}}}}
\makeatother

\newcommand{\be}{\begin{equation}}
\newcommand{\ee}{\end{equation}}
\newcommand{\bea}{\begin{eqnarray}}
\newcommand{\eea}{\end{eqnarray}}
\newcommand{\Eq}[1]{Eq.~(\ref{#1})}

\newcommand{\Fig}[1]{Fig.~\ref{#1}}

\newcommand{\mPl}{m_{\rm Pl}}
\newcommand{\LL}{\mathcal{L}}

\begin{document}


\title{The Cosmological Axino Problem}

\author{Clifford Cheung}
\affiliation{Berkeley Center for Theoretical Physics, 
  University of California, Berkeley, CA 94720, USA}
\affiliation{Theoretical Physics Group, 
  Lawrence Berkeley National Laboratory, Berkeley, CA 94720, USA}

\author{Gilly Elor}
\affiliation{Berkeley Center for Theoretical Physics, 
  University of California, Berkeley, CA 94720, USA}
\affiliation{Theoretical Physics Group, 
  Lawrence Berkeley National Laboratory, Berkeley, CA 94720, USA}

\author{Lawrence J. Hall}
\affiliation{Berkeley Center for Theoretical Physics, 
  University of California, Berkeley, CA 94720, USA}
\affiliation{Theoretical Physics Group, 
  Lawrence Berkeley National Laboratory, Berkeley, CA 94720, USA}  
\affiliation{Institute for the Physics and Mathematics of the Universe, 
  University of Tokyo, Kashiwa 277-8568, Japan}

\begin{abstract}
We revisit the cosmology of the supersymmetric QCD axion, highlighting the existence of a serious cosmological axino problem that is fully analogous to the gravitino problem of overclosure via thermal production.  A general analysis implies that the QCD axino has a mass greater than or equal to that of the gravitino in the absence of unnatural fine-tuning or sequestering.  As a consequence, bounds from thermal gravitino and QCD axino production are complementary in parameter space, and together provide a quite stringent limit on the reheating temperature after inflation given by $T_R < 10^3 - 10^6$ GeV for an axion decay constant of $f_a = 10^9 - 10^{12}$ GeV. Motivated by this result, we explore the cosmology of gravitino LSP and axino NLSP at  low $T_R$ and present three realistic scenarios for dark matter.
\end{abstract}

\maketitle

\section{Introduction}

The QCD axion is an extraordinarily elegant solution to the strong CP problem which is uniquely amenable to a variety of astrophysical and laboratory probes.  Likewise, weak-scale supersymmetry offers a theoretically motivated resolution to the gauge hierarchy problem and a wealth of implications for the LHC.  In this paper, we embrace both of these theoretical proposals and explore the cosmology of the supersymmetric QCD axion.

By construction, the QCD axion couples to the gluon with a strength inversely proportional to the axion decay constant, $f_a$.  Supersymmetry then requires a corresponding coupling of the QCD axino to the gluino.  Expressed in superspace, this interaction takes the form
\bea
{\cal L} = \frac{\sqrt{2} \alpha_3}{8\pi f_a} \int d^2\theta A W^a W^a + \textrm{h.c.},
\label{eq:axcoupling}
\eea
where the superfield containing the saxion, axion, and axino is defined as
\bea A &=& (s+ia)/\sqrt{2} + \sqrt{2} \theta \tilde a + \theta^2 F_A.
\eea
In this basis $A$ shifts by an arbitrary imaginary constant under the nonlinearly realized Peccei-Quinn (PQ) symmetry.  In this basis the axino is defined to be a PQ singlet. While  \Eq{eq:axcoupling} may be the only axino coupling present, as in the case of the KSVZ \cite{KSVZ} axion,  there can be additional couplings between the axino and other superpartners, as in the case of the DFSZ \cite{DFSZ} axion.

Since the cosmology of these theories depends crucially on the size of the axino mass, it is of utmost importance to ascertain its typical value.  On general grounds, one expects supersymmetry breaking to seep into the PQ sector by way of higher-dimensional operators that couple the axion directly to the  supersymmetry breaking field,
\bea
X &=& x + \sqrt{2}\theta \eta +\theta^2 F,
\label{eq:Xdef}
\eea   
where $F$ is the supersymmetry breaking scale.
While these contributions may be present in the form of ``Planck slop'' operators induced by unspecified ultraviolet physics, they also arise irreducibly from the calculable dynamics of supergravity \cite{Wess:1992cp}. 
Since these operators transfer supersymmetry breaking effects of order $ m_{3/2} \sim F / \mPl$ into the PQ sector, the axion and the axino acquire a mass difference of this order.  In turn, because the axion is ensured to be massless by the nonlinearly realized PQ symmetry, the axino has a mass of order $m_{3/2}$ due to Planck-scale corrections.

Concretely, this simple physical argument can be understood by the existence of a higher-dimensional operator in the effective field theory below $f_a$\footnote{Recently a similar effective theory has been considered in \cite{Higaki:2011bz}.},
\bea
 \int d^4\theta \; \frac{  (A+A^\dagger)^2 (X+X^\dagger)}{\mPl} &\sim& \frac{1}{2} m_{3/2} \tilde a \tilde a  + \ldots,
\label{eq:axmass}
\eea
which exactly preserves the full PQ symmetry,\footnote{Explicit PQ breaking operators can very easily destabilize the axion solution to the strong CP problem, even if they are generated by Planck scale dynamics \cite{Kamionkowski:1992mf,Kallosh:1995hi,Holman:1992us}.  See \cite{Georgi:1981pu,Babu:2002ic,Dias:2002hz,Cheung:2010hk} for possible resolutions to this difficulty. } and is expected from Planck-scale dynamics. Hence, the axino mass has a lower bound of order
\bea
m_{\tilde a} &\gtrsim &m_{3/2},
\label{eq:axinomass}
\eea
which provides a theoretical motivation for considering theories of low-scale supersymmetry breaking with gravitino LSP and axino NLSP.  
 \Eq{eq:axinomass} embodies a central claim of this paper: barring fine-tuning, {\it 
the axino mass acquires an irreducible contribution of order the gravitino mass which is allowed by all symmetries and is generated by uncontrolled Planck-scale physics.}  

A notable exception to this simple argument arises in certain scenarios in which extra dimensions are used to effectively sequester supersymmetric breaking from the PQ sector \cite{Randall:1998uk,Giudice:1998xp,Abe:2001cg,Choi:2009qd}.  Nevertheless, even in these theories, supersymmetry breaking is still mediated through supergravity effects which typically yield the same result as in \Eq{eq:axmass} \cite{Chun:1995hc}. In this paper we will assume that these more complicated dynamics are not at play.

\begin{figure}[t]
\begin{center} 
\includegraphics[scale=0.8]{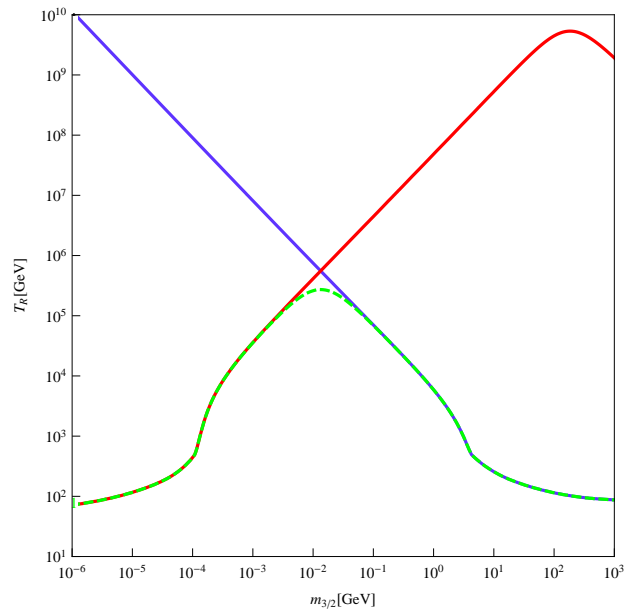}
\end{center}
\caption{The red contour is $\Omega_{3/2} h^2 = 0.11$ from gravitino production alone, and the blue contour is $\Omega_{\tilde a} h^2 = 0.11$ for axino production alone with $m_{\tilde{a}} = m_{3/2}$  and $f_a = 10^{12}$ GeV.  For larger values of $m_{\tilde{a}}$ and smaller values of $f_a$,  the blue contour is translated to the left.  The green contour can be interpreted in two ways:  (i) as the $(\Omega_{3/2}+\Omega_{\tilde a}) h^2 = 0.11$ contour for combined gravitino + axino co-dark matter when both are cosmologically stable for $m_{\tilde{a}} = m_{3/2}$;  (ii) as the $\Omega_{3/2} h^2 = 0.11$ contour for gravitino dark matter when the axino is unstable and decays sufficiently quickly to gravitinos, for any value of $m_{\tilde{a}} > m_{3/2}$.  Observe that a high reheating temperature, $T_R >3 \times 10^5 \textrm{ GeV}$, is unambiguously excluded in both cases.   The superpartner spectrum is taken to be $\{ m_{\tilde{b}},m_{\tilde{w}},m_{\tilde{g}}\} = \{100 \textrm{ GeV}, 210\textrm{ GeV}, 638\textrm{ GeV} \}$  and universal scalar masses at the GUT scale equal to $ 500 \textrm{ GeV}$.}
\label{fig:gravaxino}
\end{figure}

As we will see, the bound in \Eq{eq:axinomass} has an enormous effect on early universe cosmology.  In particular,  it implies the existence of a cosmological axino problem which is similar to the well-known gravitino problem \cite{Moroi:1993mb} but which occurs in a complementary region of $m_{3/2}$.\footnote{As in the case of the gravitino problem, this axino problem can be evaded if R-parity is broken.}  Together, the combined bounds from axino and gravitino production completely exclude the possibility of a high $T_R$, as shown in Fig.~\ref{fig:gravaxino}.  In turn, this rather unequivocally nullifies the viability of high-scale leptogenesis \cite{leptoTR} while still permitting  low-scale models of soft leptogenesis \cite{Grossman:2004dz} and testable theories of asymmetric freeze-in \cite{Hall:2010jx,Cheung:2011ph}.

Let us now briefly outline the remainder of this paper.  In Section 2, we present the theoretical rationale for \Eq{eq:axinomass}, both in general and for a canonical supersymmetric axion model.  We then go on to explore the thermal production of axinos in the early universe in Section 3.  Here we take note of a novel regime in which the dominant mode of axino production arises from the decays of superpartners still in thermal equilibrium, i.e.~freeze-in \cite{Hall:2009bx}.  Freeze-in of gravitinos and more generally of hidden sector dark matter was considered in \cite{GFI} and \cite{Cheung:2010gj,Cheung:2010gk}.  By combining the bounds on overclosure from gravitino and axino production, we can then precisely  quantify the seriousness of the cosmological axino problem.  We go on in Section 4 to consider the mixed cosmology of gravitinos, axinos, and saxions.  Applying bounds from overclosure, structure formation, and big bang nucleosynthesis (BBN), we determine the values of $m_{3/2}$, $m_{\tilde a}$ and $f_a$ for which the cosmological history is viable.  Finally, we conclude in Section 5.

\section{Supersymmetric Axion Theories}

To begin, let us provide a more rigorous justification for the naturalness argument that implies \Eq{eq:axinomass}.  
Every supersymmetric axion theory can be described by an ensemble of interacting fields
\bea
\Phi_i &=& \phi_i + \sqrt{2} \theta \psi_i + \theta^2 F_i,
\eea whose dynamics induce vacuum expectation values, $\langle \Phi_i \rangle = f_i$.  In the supersymmetric limit, the saxion, axion and axino are linear combinations of $\textrm{Re } \phi_i$, $\textrm{Im } \phi_i$ and $\psi_i$, respectively, and are massless as a consequence of the nonlinearly realized PQ symmetry.  The entirety of our discussion will occur in the language of linear fields, $\Phi_i$, so let us briefly note that the field $A$ discussed in the introduction is a linear combination of the fields $A_i = s_i + i a_i + \sqrt{2} \theta \tilde a_i + \theta^2 F_{A_i}$ where $\Phi_i = f_i e^{A_i/f_i}$.\footnote{Naively, the $\theta^2$ component of $\Phi_i$  contains a term quadratic in the axino which can produce what appears to be an axino mass in interactions involving only a single $\Phi_i$.  However, this quadratic term appears in the combination $ (F_{A_i} - \tilde a_i^2 /2 )$, and so  can always be removed by a shift of the auxiliary field.}

Next, let us consider the effect of higher-dimensional operators of the form
\be
\int d^4\theta \; \frac{\lambda_i \Phi_i^\dagger  \Phi_i (X+ X^\dagger) }{\Lambda}  =   \frac{\lambda_i f_i}{\Lambda} (F_i^\dagger F + F^\dagger F_i ) + \ldots ,  
\label{eq:UVconnector}
\ee
which couple supersymmetry breaking to the PQ sector.  Here $\lambda_i$ is an order unity dimensionless coupling and $X$ is as defined in \Eq{eq:Xdef}.   The parameter $\Lambda$ is the mass scale of the heavy particles which couple the PQ and supersymmetry breaking sectors and, as discussed in the introduction, it is at most the Planck scale,  $\Lambda \lesssim \mPl$. 

It is clear that the only fermion mass terms generated by this operator are a mixing between $\psi_i$ and $\eta$, the fermionic component of $X$. Morally, this can be understood as a rather surprising mixing term between the axino and the goldstino.  Such a mixing is strange, but by including a dynamical supersymmetry breaking sector for $X$ and diagonalizing the fermion mass matrix in the presence of \Eq{eq:UVconnector}, one goes to mass eigenstate basis and expectedly finds a massless goldstino and a heavy axino with a mass of order
\bea
m_{\tilde{a}} &\sim& \frac{F}{\Lambda},
\label{eq:maxino}
\eea
as shown explicitly for the very simple theory described in Appendix A.

One can reach the very same conclusion through a less technical, more physical argument.  In particular, the right-hand side of \Eq{eq:UVconnector} manifestly induces a non-zero value for the auxiliary field $F_i$, 
\bea
F_i &=& -\frac{\lambda_i f_i F}{\Lambda} + \ldots ,
\label{eq:Fi}
\eea
which mediates supersymmetry breaking effects proportional to $F_i$ directly into the scalar potential of the PQ sector.  Inserting appropriate powers of the characteristic scale of the PQ sector, $f_i$, we find that the mass scales and vacuum expectation values in the PQ sector shift by an amount which is of order $ F / \Lambda$.
Hence, one should expect a mass splitting between the axion and the axino of order $F/ \Lambda$ which implies \Eq{eq:axinomass}.

This argument can be restated in another way.  In particular, apply a field redefinition
\bea \Phi_i &\rightarrow& \Phi_i \left(1+ \frac{\lambda_i X}{ \Lambda} \right),
\label{eq:redef}
\eea 
which removes the connector interaction  in \Eq{eq:UVconnector} at leading order in $1/\Lambda$ at the expense of introducing $X$ directly into the PQ sector dynamics.    The effect of $X$ will be to explicitly generate an axino mass.

For example, consider the canonical supersymmetric axion theory defined by canonical Kahler potential and a superpotential
\bea
W &=& \kappa \, \Phi_3 (\Phi_1 \Phi_2 -f^2),
\eea
where $\kappa$ is a dimensionless coupling, and a straightforward calculation shows that the vacuum is stabilized at 
$ f_1 f_2 = f^2$ and $f_3 = 0$ in the supersymmetric limit.\footnote{Our results hold irrespective of the precise dynamics which break supersymmetry---indeed, one obtains the correct answer even when treating $X$ as a non-dynamical spurion of supersymmetry breaking.}  Consider the effect of  the higher-dimensional operators in \Eq{eq:UVconnector} for the symmetrical case $\lambda_{1,2} = \lambda$ and $\lambda_3 =0$.  After the field redefinition in \Eq{eq:redef}, a simple calculation shows that the vacuum is slightly shifted so that $f_3 =- \lambda F / \kappa \Lambda$.  This induces a mass for the axino equal to $\lambda F / 2 \Lambda$ from the superpotential, which again accords with \Eq{eq:axinomass}.

Before we continue on to cosmology, let us comment briefly on some of the existing literature on the mass of the axino.  While statements are frequently made to the effect that the axino mass is highly model dependent, we disagree.  The authors of \cite{Chun:1995hc} and \cite{Goto1992103} computed the axino mass in a variety of settings and verified \Eq{eq:axinomass} and \Eq{eq:maxino} in all cases. The only exceptions resulted from the imposition of special relations among disparate and unrelated parameters to yield a lighter axino. 
We conclude that in generic theories the axino mass is given by \Eq{eq:maxino} and the only model dependence is the value of $\Lambda$, which has an upper bound of $\mPl$.

\section{Axino Cosmology}

Given a proper theoretical justification of \Eq{eq:axinomass}, 
let us now consider the early universe cosmology of the axino.  Like the gravitino, the axino is produced through thermal scattering and decay processes.
As observed by Strumia \cite{Strumia:2010aa}, one can trivially compute axino production by translating every equation relevant to gravitino production by the simple replacement
\bea
\frac{m_{\tilde g}}{F} &\leftrightarrow& \frac{\sqrt{2}\alpha_3}{4\pi f_a},
\eea
while shutting off all scattering and decay processes involving gauginos and scalars not accounted for in \Eq{eq:axcoupling}.

\begin{figure}[t]
\begin{center} 
\includegraphics[scale=0.8]{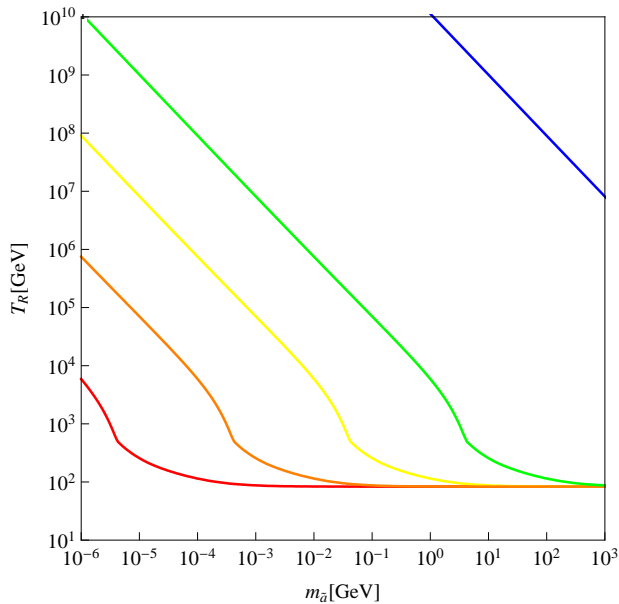}
\end{center}
\caption{Contours of $\Omega_{\tilde a} h^2 = 0.11$ for the KSVZ \cite{KSVZ} model, where axinos couple minimally, for $m_{\tilde{g}} = 638 \textrm{ GeV}$.  
The $\{$red, orange, yellow, green, blue$\}$ contours correspond to 
$\log_{10} f_a/\textrm{GeV} = \{9,10,11,12,15\}$.}
\label{fig:axino}
\end{figure}

\begin{figure}[t]
\begin{center} 
\includegraphics[scale=0.8]{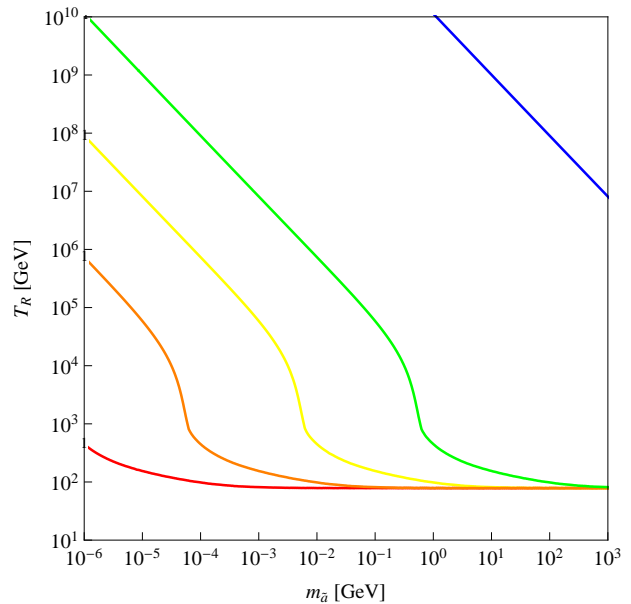}
\end{center}
\caption{Contours of $\Omega_{\tilde a} h^2 = 0.11$ for the DFSZ \cite{DFSZ} model, where axinos have additional couplings to squarks, for $m_{\tilde{g}}= 638$ GeV and $m_{\tilde{q}} = 500 \textrm{ GeV}$. The new decay $\tilde{q} \rightarrow \tilde{a} + q$, results in a larger freeze-in region, where the contours are nearly vertical, as compared to \Fig{fig:axino}.   The  contour colors correspond to the same values of $f_{a}$ as in \Fig{fig:axino}.}
\label{fig:DFSZaxino}
\end{figure}

In \Fig{fig:axino} and \Fig{fig:DFSZaxino}, we have numerically plotted contours of $\Omega_{\tilde a} h^2 = 0.11$ for $m_{\tilde g} = 638\textrm{ GeV}$ and different values for the axion decay constant, assuming the axino to be cosmologically stable.  As one can see, for axion decay constants within the ``axion window'',  $10^9 \textrm{ GeV} < f_a < 10^{12} \textrm{ GeV}$, the bounds on $T_R$ are quite stringent.  For a GUT scale axion decay constant corresponding to the ``anthropic window'', the bound is weak.   Moreover, as in the case of gravitino cosmology, one discovers regions where axino production arises dominantly from freeze-in and scattering, corresponding to the vertical and sloped portions of the contours, respectively.  Note that the axino contours in \Fig{fig:axino} are ``flipped'' from the usual contours corresponding to gravitino overclosure, simply because the scattering and decay rates for axinos are  independent of the axino mass and fixed by $f_a$.

\subsection{Cosmological Axino Problem}

The cosmology of gravitinos and axinos has been studied in great detail by numerous authors over many years \cite{axinoGravCosm}.  Nevertheless, to our knowledge the existence of a serious cosmological axino problem has never been clearly discussed. In particular, the red and blue curves in \Fig{fig:gravaxino} are bounds on $T_R$ from gravitino and axino production overlayed, assuming that they are co-dark matter candidates and $m_{\tilde a} = m_{3/2}$.  This is a conservative assumption to make because as shown in \Eq{eq:axmass}, the axino acquires an irreducible contribution to its mass of order the gravitino mass which cannot be eliminated without  arbitrary fine-tuning.  On the other hand, if for whatever reason the axino is heavier than the gravitino mass, then this only harshens axino bounds depicted in \Fig{fig:gravaxino}, since the axino contour is translated to the left by a factor $m_{\tilde{a}}/m_{3/2}$.

\Fig{fig:gravaxino} illustrates a primary claim of this paper: {\it the combined gravitino and axino problems entirely exclude the possibility of a high reheating temperature.}  The upper bound on $T_R$ is about $3\times 10^{5}$ GeV for $f_a = 10^{12}$ GeV, and is even lower for lower $f_a$.  Thus, supersymmetry plus the axion solution to the strong CP problem are strongly at odds with theories of high $T_R$ and high-scale leptogenesis, which typically require $T_R > 10^9 \textrm{ GeV}$.  The bound on $T_R$ is greatly relaxed for very large $f_a$, as could occur with anthropic selection of a small axion misalignment angle.   


\begin{figure}[t]
\begin{center} 
\includegraphics[scale=0.8]{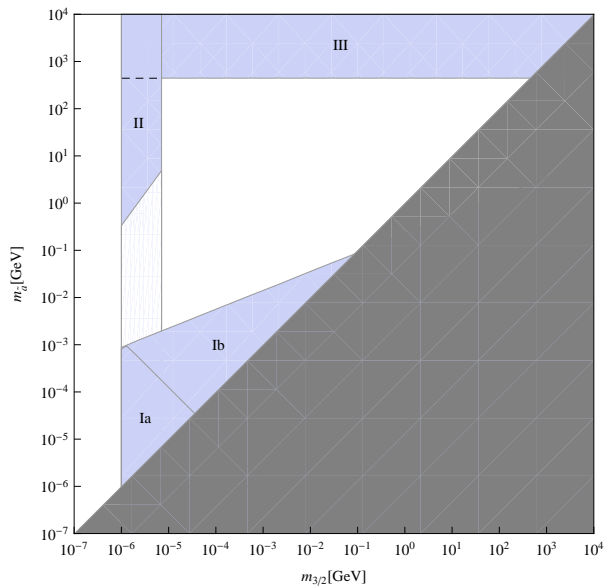}
\end{center}
\caption{Light blue regions in the $(m_{3/2}, m_{\tilde{a}})$ plane have realistic dark matter cosmologies with a gravitino LSP and an axino NLSP for $f_{a} = 10^9$ GeV.   Regions Ia and Ib correspond dark matter comprised of both stable gravitinos and axinos, with gravitinos or axinos dominating the dark matter abundance, respectively. Region II is bounded by warm dark matter constraints and limits on relativistic species. In Region III the axino mass is sufficiently large that the produced gravitinos are cold.  The superpartner spectrum is taken to be identical to that of \Fig{fig:gravaxino}.}
\label{fig:axbound9}
\end{figure}

\begin{figure}[t]
\begin{center} 
\includegraphics[scale=0.8]{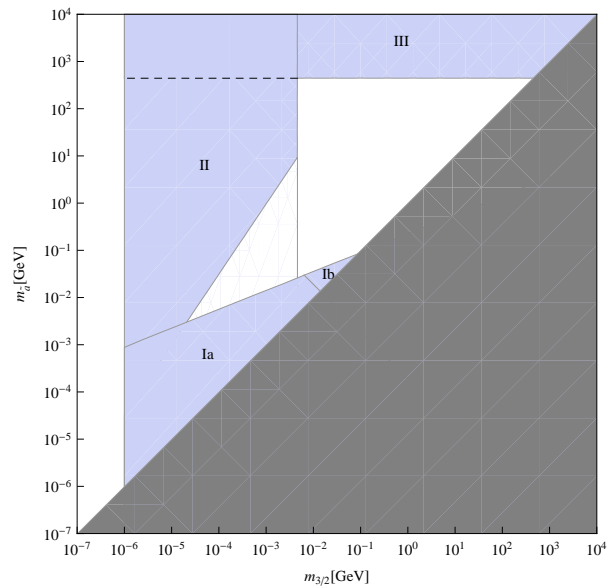}
\end{center}
\caption{Same as \Fig{fig:axbound9} but with $f_{a} = 10^{12}$ GeV. Region II, corresponding to sub-dominant hot gravitino dark matter from axino decay, has increased. The superpartner spectrum is taken to be identical to that of \Fig{fig:gravaxino}.}
\label{fig:axbound12}
\end{figure}

\section{Gravitino and Axino Cosmology}
\label{sec:gravaxcosmo}

Finally, let us consider cosmology in theories with gravitino LSP and axino NLSP, keeping careful track of the decays of axinos to gravitinos and cosmological limits on both axinos and saxions.  The  operator in \Eq{eq:axmass} generates an axino-axion-gravitino coupling of the form
\bea
\frac{m_{\tilde{a}}}{m_{3/2}\mPl} \tilde{G} \sigma^{\mu} \tilde{a}^\dagger \partial_{\mu}a,
\label{eq:gaa}
\eea
where $\tilde{G}$ is the physical goldstino.  Note that this operator respects the PQ symmetry because the axion is derivatively coupled.   The above operator mediates the decay of the axino to the gravitino and axion with a lifetime
\bea
\tau_{\tilde{a}} \simeq 10^{9} \textrm{ sec} \times \left( \frac{m_{3/2}}{\mbox{MeV}} \right)^2 
\left( \frac{\mbox{GeV}}{m_{\tilde{a}}} \right)^5.
\label{eq:axinoLifetime}
\eea
This lifetime ranges over many orders of magnitude in the $\left(m_{3/2},m_{\tilde{a}}\right)$ plane, yielding a broad spectrum of cosmological histories which we now explore.

\subsection{Three Scenarios for Dark Matter}

The cosmological history varies substantially in the $\left(m_{\tilde{a}},m_{3/2}\right)$ plane, as depicted in \Fig{fig:axbound9} and \Fig{fig:axbound12} for $f_{a}=10^{9}$ GeV and $10^{12}$ GeV respectively.  The grey region corresponds to $m_{\tilde a} < m_{3/2}$, which is disfavored by the naturalness argument leading to \Eq{eq:axmass}, while the blue regions labeled by Ia, Ib, II, and III are consistent with existing experiments.

Region I of these figures, $\tau_{\tilde{a}} \gtrsim 10^{18}$ sec, corresponds to a ``Cosmologically-Stable Axino". In this regime, gravitinos and axinos are co-dark matter with $m_{\tilde{a}}Y_{\tilde{a}} + m_{3/2}Y_{3/2} = \rho_0 /s_0$, where
\bea
\frac{\rho_0}{s_0} &=&  4 \times 10^{-10} \textrm{ GeV}
\eea
is the observed dark matter relic abundance today.  Here $Y_{\tilde a}$ and $Y_{3/2}$ are the thermal axino and gravitino yields arising from scattering and decay processes. Provided that $m_{\tilde{a}}, m_{3/2} >$ keV, both components are cold.   Region Ia corresponds to the dark matter energy density dominated by gravitinos, while Region Ib corresponds to axino domination.

Meanwhile, as the axino mass increases, the lifetime for the decay $\tilde{a} \rightarrow \tilde{G} + a$ decreases, so the axinos become cosmologically unstable. Gravitinos from axino decay will then contribute to the total gravitino dark matter abundance, requiring  $m_{3/2}\left(Y_{\tilde{a}} + Y_{3/2}\right) = \rho_0/s_0$.

If  $m_{\tilde{a}} \gg m_{3/2}$ the gravitinos produced in axino decay are highly relativistic, becoming non-relativistic only at a temperature
\bea
T_{\textrm{NR}} \sim 10^{-2}  \textrm{ eV} \times \left( \frac{m_{\tilde{a}}}{\mbox{GeV}} \right)^{3/2}.
\label{eq:TNR}
\eea
In Region III of \Fig{fig:axbound9} and \Fig{fig:axbound12}, the axino mass $m_{\tilde{a}} \gtrsim \textrm{TeV}$, so that $T_{\textrm{NR}} \gtrsim$ keV and the gravitinos from axino decay are cold. There is little parameter space for this ``Fast Decaying Axino" case, since axinos are are no-longer expected to be the NLSP if $m_{\tilde{a}}$ increases much above a TeV. 

For cosmologically unstable axinos,  $m_{\tilde{a}} \lesssim \textrm{TeV}$ gives $T_{\textrm{NR}} \lesssim$ keV so that the gravitinos from axino decay are warm and will over-deplete sub-Mpc structures unless they constitute less than 3\% of dark matter \cite{WDM}. The right-hand vertical contour that bounds Region II,  the ``Sub-Dominant Axino" case, corresponds to $Y_{\tilde{a}} < 0.03  Y_{3/2}$, placing an upper bound on $m_{3/2}$ which increases with $f_{a}$
as seen by comparing \Fig{fig:axbound9} and \Fig{fig:axbound12}.  

The lower bound of Region II arises from the adverse effects of relativistic gravitinos arising from axino decay.  As discussed in detail in  \cite{Kawasaki:2007mk}, additional relativistic degrees of freedom can adversely affect matter-radiation equality and are thus  constrained by observations of the cosmic microwave background, galaxy clustering, and the Lyman-$\alpha$ forest.  These bounds apply to axino lifetimes shorter than $10^{13}$ sec, and correspond roughly to an additional effective number of neutrino species less than one.  Numerically, this implies \cite{Kawasaki:2007mk}
\bea
m_{\tilde a} Y_{\tilde a} \lesssim 3.4\times 10^{-5} \textrm{ GeV} \times \left(\frac{10}{g_{*s}(T_{\tilde a})}\right) \left( \frac{T_{\tilde a}}{\rm MeV}\right),
\label{eq:KawBound}
\eea
where $T_{\tilde a}$ is the temperature when the axino decays.  Here $T_{\tilde a}$ is simply computed by solving for the temperature at which $H = 1/\tau_{\tilde a}$, where $\tau_{\tilde a}$ is defined in \Eq{eq:axinoLifetime}.  Plugging in the value of $Y_{\tilde a}$ from scattering in \Eq{eq:KawBound} then yields a substantial limit on $T_R$ given by
\bea
T_R &<& 200 \textrm{ GeV} \times \left(\frac{f_a}{10^{12} \textrm{ GeV}} \right)^2 \left(\frac{m_{\tilde a}}{ \textrm{GeV}} \right)^{3/2} \left(\frac{\rm GeV}{m_{3/2}} \right)\nonumber
\eea
For $m_{\tilde a} \sim m_{3/2}$ the above bound on $T_R$ is much more stringent than the one depicted in 
\Fig{fig:gravaxino} from overclosure from gravitinos and axinos.  However, by fixing $T_R$ to lie on the green contour of \Fig{fig:gravaxino}, such that gravitino dark matter from thermal production equals the measured value today, then the above inequality for $T_R$ can be reinterpreted as a lower bound on $m_{\tilde a}$ as a function of $m_{3/2}$.  This constraint is what produces the slanted lower-right boundary of Region II.  

\begin{figure}[t]
\begin{center} 
\includegraphics[scale=0.8]{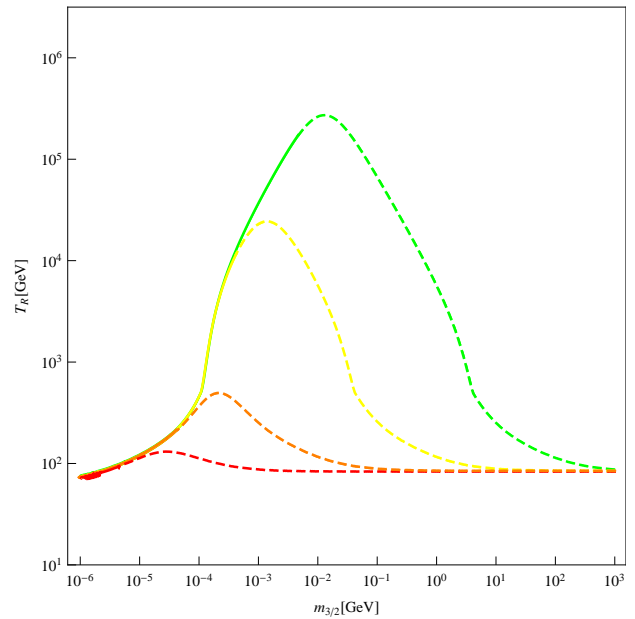}
\end{center}
\caption{Gravitino dark matter in Region II, with contours of $m_{3/2}\left(Y_{\tilde{a}}+Y_{3/2}\right) = \rho_0/s_0$. Here the $\{$Red, Orange, Yellow, Green$\}$ contours correspond to $\log_{10} f_{a}/\textrm{GeV}=\{9,10,11,12\}$. The dotted portion of each contour shows values of $m_{3/2}$ that are disallowed by hot dark matter, where hot gravitinos from axino decay constitute more than 3\% of the total  abundance. The transition from solid to dashed contour corresponds to crossing from Region II into the white region of \Fig{fig:axbound9} and \Fig{fig:axbound12}. The superpartner spectrum is taken to be identical to that of \Fig{fig:gravaxino}.}
\label{fig:changefa}
\end{figure}

 
In \Fig{fig:changefa} contours of $\Omega_{3/2} h^2 = 0.11$ have been plotted for several values of $f_{a}$ and a fixed superpartner spectrum. These are contours of $m_{3/2}\left(Y_{\tilde{a}}+Y_{3/2}\right) = \rho_0/s_0$ and therefore corresponds to gravitino dark matter in the ``Sub-Dominant Axino" case. The contours become dashed when the values of $m_{3/2}$ are excluded by the warm dark matter bound  $Y_{\tilde{a}}<0.03Y_{3/2}$; this corresponds to the rightmost boundary of Region II in \Fig{fig:axbound9} and \Fig{fig:axbound12}. Examining  \Fig{fig:changefa} it is clear that while the angled gravitino scattering region is further excluded, the vertical gravitino freeze-in region is relatively unaffected. For $f_{a}=10^{11}$ GeV and $10^{12}$ GeV the axino and gravitino yields in the bound $Y_{\tilde{a}}<0.03Y_{3/2}$ are both dominated by the scattering contributions. Thus the $T_{R}$ dependence cancels and the resulting upper bound on $m_{3/2}$ is dependent only on $f_a$. As $f_{a}$ decreases the $m_{3/2}Y_{\tilde{a}}$ contours move to left as seen in \Fig{fig:axino}. Therefore the gravitino yield that is to be compared with the axino scattering yield becomes dominated by freeze-in, forcing $T_{R}$ to smaller values. Since  $Y_{\tilde{a}}^{\textrm{scatt}}/Y_{3/2}^{\textrm{decay}} \propto m_{3/2}^2 T_{R} / f_{a}^{2}$ the upper bound on $m_{3/2}$ relaxes whenever we cross over into the freeze-in dominated region.

 \Fig{fig:axbound9} and \Fig{fig:axbound12} depict cosmological restrictions on $(m_{3/2}, m_{\tilde a}, f_a)$ which limit the range of collider signals resulting from the decay of the lightest observable-sector superpartner (LOSP)---that is, the lightest R-parity odd superpartner of a standard model particle.  Assuming an axion model in which the LOSP couples directly to the axino (for instance a gluino LOSP in the KSVZ model) the ratio of LOSP decay rates to axinos and gravitinos is proportional to $(m_{3/2}/f_a)^2$, then branching ratio to axinos (gravitinos) dominates at large (small) $m_{3/2}$\footnote{Regardless of the choice of axion model any ultraviolet physics which generates the operator in \Eq{eq:axmass} will typically induce axino/goldstino kinetic mixing via  $\int d^4\theta \; \epsilon (A+A^\dagger) (X+X^\dagger) $, where $\epsilon \sim f_a/\mPl$.  In components, $\epsilon \tilde{G} \sigma^{\mu} \partial_{\mu} \tilde{a}^\dagger$ is removed by shifting $\tilde{G} \rightarrow \tilde{G} + \epsilon \tilde{a}$, which induces a coupling of the axino to the supercurrent of the form $\epsilon\tilde{a} \tilde{J}$. Therefore, the LOSP typically decays to $\tilde{a}$ with a branching fraction suppressed by a factor of $\epsilon^2$ relative to the branching fraction to $\tilde{G}$.}.   Hence in Region III at large $m_{3/2}$ the LOSP lifetime is fixed by $f_a$ and the dark matter is  produced via axino production.  At some intermediate values of $m_{3/2}$ the LOSP has sizable branching ratios to both axinos and gravitinos, and if the axino mass is sufficiently heavy these modes can be distinguished by kinematics.   Hence in parts of regions I and II it may be possible to measure
$(m_{3/2}, m_{\tilde a}, f_a)$.  At a particular value of $m_{3/2}$, for instance around $m_{3/2} \sim 0.3$ MeV for squark masses of 500 GeV, the dark matter is produced by gravitino freeze-in.  This value of $m_{3/2}$ is sufficiently small that the LOSP decays dominantly to gravitinos and has a lifetime directly correlated with the freeze-in production mechanism \cite{GFI}.  For the charged slepton LOSP, decay signals to gravitinos and axinos have been studied in \cite{Brandenburg:2005he}, examining the degree to which axino and gravitino modes can be distinguished.

\subsection{Cosmological Bounds from Saxions}

So far we have ignored the saxion component of the axion supermultiplet.   
With communication between the supersymmetry breaking and PQ sectors mediated at mass scale $\Lambda$, the saxion picks up a mass at the same order as the axino mass, $m_s \sim F/\Lambda$.   This follows from very similar arguments to those of Section 2---in particular,  by inserting one power of $F_i$ of \Eq{eq:Fi} into the operator of \Eq{eq:UVconnector}, or more directly from the dimension 6 operator $\Phi_i^\dagger \Phi_i X^\dagger X/\Lambda^2$ in the Kahler potential.   In the effective theory beneath  $f_a$,  Planck-scale dynamics induce the operator
\bea
\int d^4\theta \; \frac{ X^\dagger X (A^\dagger+A)^2}{\mPl^2} &\simeq& \frac{1}{2}m_{3/2}^2 s s + \textrm{h.c.} + \ldots,
\label{eq:sxmass}
\eea
so that the gravitino mass provides a lower bound for the saxion mass.   

The cosmological bounds imposed by saxions have been studied in detail in \cite{Kawasaki:2007mk}.  We find that under the assumption $m_{s} > m_{3/2}$ these bounds on $T_R$ are typically less stringent than those coming from axino cosmology.   In particular supersymmetric axion models must satisfy constraints on
\begin{itemize}
\item{Saxion overclosure,}
\item{Saxion decays influencing BBN,}
\item{Relativistic degrees of freedom.}
\end{itemize}
The saxion abundance receives a contribution from scattering in the early universe which is identical to the axino scattering yield $Y_{s}^{\textrm{scatt}} = Y_{\tilde{a}}^{\textrm{scatt}}$, as can be seen from the supersymmetric interaction of \Eq{eq:axcoupling}.  We assume a negligible contribution to the saxion abundance from coherent oscillations. Unlike the axino and gravitino, saxion overclosure does not pose a threat because saxions decay rapidly; to photons through the electromagnetic analogue of \Eq{eq:axcoupling} and to axions through the kinetic term of the axion supermultiplet.  

Since the saxion will rapidly decay to photons one may worry that these photons might ruin the predictions of BBN unless the saxion is made heavy enough. However, this is not the case, since decays to photons are suppressed by a loop factor and so will always be subdominant to decays to axions
\bea
\frac{\Gamma\left( s \rightarrow aa\right)} {\Gamma \left(s \rightarrow \gamma\gamma\right)} \sim \frac{64 \pi^2}{\alpha^2},
\label{eq:fracGamma}
\eea
where $\alpha$ is the fine structure constant.  While the branching fraction of the saxion to axions can in principle be small, this requires a delicate and unnatural cancellation in the underlying theory.


Finally, let us consider the issue of relativistic degrees of freedom.
In complete analogy with the axino, the saxion will decay and produce relativistic energy in axions which is subject to the bound in \Eq{eq:KawBound} taken from \cite{Kawasaki:2007mk} except with $m_{\tilde a} Y_{\tilde a}$ and $T_{\tilde a}$ replaced with $m_s Y_s$ and $T_s$.  Plugging in for $Y_s$, the saxion yield from thermal scattering (which is essentially equal to the axino yield from thermal scattering), and plugging in for $T_s$, the temperature at the time of saxion decay where  $\tau_s^{-1} \sim m_s^3 /64 \pi f_a^2$, one acquires a bound on $T_R$ given by
\bea
T_R &<& 5\times 10^8 \textrm{ GeV} \times \left(\frac{f_a}{10^{12} \textrm{ GeV}} \right) \left(\frac{m_s}{ \textrm{GeV}} \right)^{1/2}, \nonumber
\eea
which is generally weaker than the $T_R$ bound from overclosure depicted in \Fig{fig:gravaxino}.


\section{Conclusions}

If supersymmetry and the QCD axion coexist, then a large range of parameter space will include a gravitino LSP and axino NLSP.   In this case the axino plays an important role in determining viable cosmological histories.  In particular, the overclosure constraint alone provides a very powerful limit on the reheat temperature, as shown in Figure 6, that can be approximated by
\bea
T_R < 3 \times 10^5 \, \mbox{GeV} \, \left( \frac{f_a}{10^{12} \, \mbox{GeV}} \right).
\label{eq:TRbound}
\eea
 We identify three very different phases of combined gravitino and axino cosmology which we label according to the nature of the axino:
 
 \begin{list}{\labelitemi}{\leftmargin=2.5em}
\item[{ (I)}]
{\bf ``Cosmologically-Stable Axino"} is the case where axinos and gravitinos are co-dark matter.
\item[{ (II)}]
{\bf ``Sub-Dominant Axino"} production leads to gravitino dark matter from decays, with an upper bound on $m_{3/2}$ that depends on $f_a$.
\item[{ (III)}]
{\bf ``Fast-Decaying Axino"} has a TeV scale mass and thus decays sufficiently early that  gravitino dark matter is cold.
\end{list}


The regions of gravitino and axino masses that yield these three cosmologies are shown in \Fig{fig:axbound9} and \Fig{fig:axbound12} for $f_a = 10^9$ GeV and $10^{12}$ GeV.  In all three cosmologies, dark matter may be dominantly produced by gravitino freeze-in.

The required ranges of $T_R$ for cases (II) and (III) are shown in \Fig{fig:changefa} for several values of $f_a$. They give an upper bound on $T_R$ that is much lower than for the case of gravitino dark matter without axinos, and is also lower than the bound of \Eq{eq:TRbound}.  Thus the presence of the axinos increases the likelihood that the production mechanism for gravitino dark matter is freeze-in from decays, strengthening the possibility that LHC will provide strong evidence for gravitino dark matter by measuring the LOSP lifetime \cite{GFI}.     Finally, let us note that our  conclusions are unaltered by cosmological considerations of the saxion for any value of its mass greater than $m_{3/2}$.

\begin{acknowledgements}
We thank Mina Arvanitaki, Kiwoon Choi, and Peter Graham for useful discussions. This work was supported in part by the Director, Office of Science, Office of High Energy and Nuclear Physics, of the US Department of Energy under Contract DE-AC02-05CH11231 and by the National Science Foundation on grant PHY-0457315. 
\end{acknowledgements}

\appendix

\section{A Simple Theory}

Consider a minimal theory of PQ and supersymmetry breaking defined by
\bea
\LL &=& \int d^4\theta \; K + \int d^2\theta \; W + {\rm h.c.} ,
\eea
where the Kahler potential and and superpotential are defined by 
\bea
K &=& G(\Phi^\dagger \Phi) + H(X^\dagger, X) +\frac{\Phi^\dagger\Phi X}{\Lambda} + {\rm h.c.} \\
W &=& F X.
\eea
Here $\Phi$ and $X$ are the PQ and supersymmetry breaking fields, respectively, and $G$ and $H$ are as of yet unspecified real functions.  Since $\Phi$ is PQ charged, it cannot be present in the superpotential.  Also, note that $G$ is purely a function of $\Phi^\dagger \Phi$ so that the PQ symmetry is left unbroken.  We interpret the $1/\Lambda$ suppressed operator coupling $\Phi$ and $X$ as something generated by unspecified high-scale dynamics.

Demanding that $G$ and $H$ have a form such that $\langle \Phi \rangle = f_a$ and $\langle X\rangle =0$ at the minimum of the scalar potential implies that
\bea
 \langle G^{(1)} \rangle &=& f_a^2  \langle G^{(2)}\rangle+ f_a^4 \langle G^{(3)}\rangle  \\ 
\langle H^{(1,2)} \rangle = \langle H^{(2,1)} \rangle  &=& -\frac{f_a^2}{\Lambda^3 ( \langle G^{(1)}\rangle  + f_a^2 \langle G^{(2)}\rangle)^2},\nonumber
\eea
where the superscripts denote derivatives with respect to the function arguments.
Plugging these expressions into the action, canonically normalizing the fermion kinetic terms, and then diagonalizing the fermion mass matrix, one discovers that one linear combination of fermions from $\Phi$ and $X$ is massless, as is expected of the goldstino.  Meanwhile, the orthogonal axino component acquires a mass
\bea
m_{\tilde a} &=& \frac{F}{2 \Lambda}  \frac{1}{\langle G^{(1)} \rangle + f_a^2 \langle G^{(2)} \rangle}\frac{1}{\langle H^{(1,1)} \rangle}+ \ldots,
\eea
where the ellipses denote terms higher order in $\Lambda$.  Hence, we confirm the result of the simple operator argument in \Eq{eq:axinomass}.

\bea
\nonumber
\eea

\end{document}